\documentclass[reprint, superscriptaddress, amsmath, amssymb, aps, prl]{revtex4-2}

\usepackage{amsmath}
\usepackage{amssymb}
\usepackage{indentfirst}
\usepackage{commath}
\usepackage{graphicx}
\usepackage[caption=false,position=bottom,labelfont={bf}]{subfig}
\usepackage{overpic}
\usepackage{dcolumn}
\usepackage{bm}

\usepackage[colorlinks,linktocpage,hypertexnames=false]{hyperref}
\usepackage{xcolor}
\definecolor{myred}{rgb}{0.8,0.1,0.2}
\definecolor{myblue}{rgb}{0.1,0.2,0.6}
\hypersetup{
    colorlinks=true,
    linktoc=all,
    linkcolor={myred},
    citecolor={myblue},
    urlcolor={myblue},
}

\newcommand{\bra}[1]{\left\langle#1\right|}
\newcommand{\ket}[1]{\left|#1\right\rangle}


\begin{document}
%--------------------------------------------------------------------------

%--------------------------------------------------------------------------
\title{
The interplay of phase fluctuations and nodal quasiparticles:\\
ubiquitous Fermi arcs in two-dimensional $\textit{d}$-wave superconductors
}
%--------------------------------------------------------------------------

\author{Xu-Cheng Wang}
\affiliation{State Key Laboratory of Surface Physics, Fudan University, Shanghai 200433, China}
\affiliation{Center for Field Theory and Particle Physics, Department of Physics, Fudan University, Shanghai 200433, China}

\author{Xiao Yan Xu}
\email{xiaoyanxu@sjtu.edu.cn}
\affiliation{Key Laboratory of Artificial Structures and Quantum Control (Ministry of Education), School of Physics and Astronomy, Shanghai Jiao Tong University, Shanghai 200240, China}
\affiliation{Hefei National Laboratory, Hefei 230088, China}

\author{Yang Qi}
\email{qiyang@fudan.edu.cn}
\affiliation{State Key Laboratory of Surface Physics, Fudan University, Shanghai 200433, China}
\affiliation{Center for Field Theory and Particle Physics, Department of Physics, Fudan University, Shanghai 200433, China}
\affiliation{Hefei National Laboratory, Hefei 230088, China}

\date{\today}

\maketitle

% {\color{myblue}\it Introduction.}
%--------------------------------------------------------------------------
The phenomenon of pseudogap and Fermi arcs has been a long-standing puzzle in the field of unconventional superconductivity (SC).
Since first discovered in underdoped cuprates~\cite{timusk_pseudogap_1999},
the pseudogap has been generally identified in various (quasi) two-dimensional (2D) contexts,
including thin-film FeSe~\cite{faeth_incoherent_2021},
layered heavy fermion systems~\cite{donovan_observation_1997},
magic angle twisted bilayer graphene~\cite{oh_evidence_2021},
and recently even ultra-cold atoms~\cite{li_observation_2024}.
One prominent explanation for the pseudogap is hence to treat it as a precursor of pairing due to phase decoherence.

In contrast, the generality of Fermi arcs among quasi-2D nodal systems is rarely observed or predicted.
Theoretical attempts relating phase fluctuations to pseudogap or Fermi arcs
trace back to Refs.~\cite{franz_phase_1998,kwon_effect_1999,berg_evolution_2007},
where their theories focus mainly on cuprates with $d$-wave symmetry,
and a concise yet universal criterion for the evolution of pseudogap and Fermi arcs is lacking.
From the numerical side, a series of studies~\cite{eckl_pair_2002,han_pseudogap_2010,singh_fermi_2022}
investigated the phase-fluctuation-driven Fermi arcs
by coupling electrons to classical pairing fields
and combining exact diagonalization with classical Monte Carlo sampling.
It is quite rare to observe Fermi arcs and, moreover, attribute their origin to phase fluctuations
within a fully quantum, correlated, and unbiased model.
As a result, although the idea was put forward for years,
an intuitive yet quantitative picture of the phase-fluctuation-driven Fermi arcs is missing,
not to mention confirming it as a convergence of independent theory and numerics.

In this study, we theoretically show that
in the normal state of general 2D nodal superconductors,
the pseudogap and Fermi arcs can universally emerge
due to thermal (static) phase fluctuations, without any competing order.
Combining perturbative theory with the disorder-average technique,
it is revealed that
the evolution of Fermi arcs is quantitatively described by two emergent characteristic length scales of the system:
one is the normal-state superconducting correlation length $\xi(T)$,
and another the nodal Bardeen-Cooper-Schrieffer (BCS) coherence length $\xi_\text{BCS}(\bm{k})$.
To support our findings, we numerically report the observation of Fermi arcs in a Hubbard-like model,
proposed by one of our co-authors~\cite{xu_competing_2021},
with sign-problem-free determinant quantum Monte Carlo (DQMC) calculations.
% As far as we noticed, it is the first time in a correlated model that
% Fermi arcs are identified with unbiased simulations.
Despite the presence of quantum fluctuations and interaction effect,
the numerical results for the pair-scattering rate $\Gamma_\text{pf}$
exhibit excellent agreements with our theoretical predictions,
where $\Gamma_\text{pf}$ is expected to scale linearly with the inverse superconducting correlation length $\xi(T)^{-1}$.
The generality of our findings suggests
that Fermi arcs could potentially be observed not only in the cuprate family,
but also in other quasi-2D nodal superconductors and ultra-cold atoms.
%--------------------------------------------------------------------------

\begin{figure*}[tp]
    \centering
    \includegraphics[width=.9\linewidth]{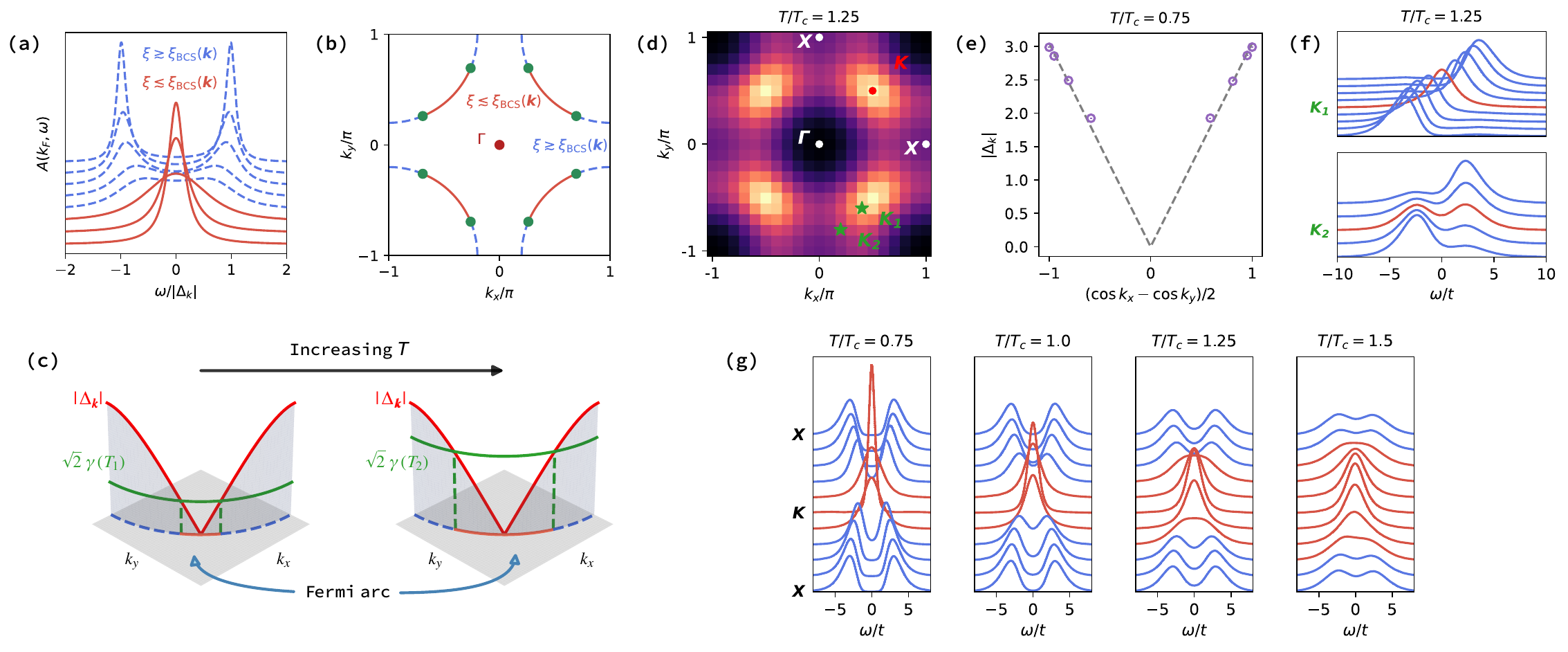}
    \caption{
        (a)-(c) Theoretical illustration of pseudogap and Fermi arcs.
        (a) Evolution of spectral weights $A(k_F,\omega)$ as a competition of $\xi$ and $\xi_\text{BCS}(k)$.
        (b) Schematic plot of the Fermi arcs.
        (c) Evolution of Fermi arcs with increasing temperature.
        (d)-(g) Observation of Fermi arcs in the Hubbard-like model with DQMC on a $L=20$ lattice.
        (d) Green's function $\beta G(\bm{k},\beta/2)$ as an estimation of $A(\bm{k},\omega=0)$ at $T/T_c=1.25$.
            The node/antinode is labeled as $K$/$X$.
        (e) $d$-wave gap function $\abs{\Delta_{\bm{k}}}$ extracted directly from $A(\bm{k},\omega)$ at $T/T_c=0.75$.
        (f) $A(\bm{k},\omega)$ along the momentum path perpendicular to Fermi surface.
            The momentum paths intersect with the Fermi surface at $K_1$ and $K_2$ respectively,
            as marked in (d) where $K_1$ belongs to the Fermi arc while $K_2$ is not.
        (g) $A(\bm{k},\omega)$ along the node-antinode line $X$-$K$-$X$ with varying $T$.
            The red curves outline the shape of Fermi arcs.
    }
    \label{fig:fig1}
\end{figure*}

% {\color{myblue}\it Phase fluctuations and Fermi arcs.}
%--------------------------------------------------------------------------
Our theory is based on a phenomenological Hamiltonian $H=H_0+V$,
where
$
    H_0 = \sum_{\sigma}\int\mathrm{d}^2\bm{r}\ \psi_\sigma^\dag(\bm{r}) (-\frac{\nabla^2}{2m}-\mu) \psi_\sigma(\bm{r})
$
and
$
    V = \int\mathrm{d}^2\bm{r} \mathrm{d}^2\bm{s}\ \Delta(\bm{r},\bm{s}) \psi_\uparrow^\dag(\bm{r}+\bm{s}/2) \psi_\downarrow^\dag(\bm{r}-\bm{s}/2) + \textrm{h.c.}
$
describes generic superconducting pairing.
The superconducting order parameters $\Delta(\bm{r},\bm{s})$ depend on the intrinsic momentum through
$\int\mathrm{d}^2\bm{s}\ \mathrm{e}^{-i\bm{p}\bm{s}}\Delta(\bm{r},\bm{s}) = \Delta(\bm{r})\varphi(\bm{p})$,
where $\varphi(\bm{p})$ denotes the pairing form factor,
e.g. $\varphi_d(\bm{p})=(p_x^2-p_y^2)/p^2$ for the $d_{x^2-y^2}$ pairing.
Throughout this work, we assume that:
(1) Due to the Berezinskii-Kosterlitz-Thouless (BKT) nature of 2D superconducting transition,
    the phase fluctuations of $\Delta(\bm{r},\bm{s})$ dominate
    while its amplitude $\abs{\Delta(\bm{r},\bm{s})}$ maintains the mean-field value $\Delta_0$,
    which is also known as the scenario of preformed pairs.
(2) For a finite-temperature transition, the temporal fluctuations are neglectable compared with the spatial ones.
    Namely, we consider here only the static fluctuations.
    % This can be argued from the fact that
    % as the BKT transition point of SC is approached from higher temperatures,
    % the spatial correlation length $\xi$ formally diverges
    % while the correlation in the imaginary-time direction is bounded by a finite inverse temperature $\beta=(k_B T)^{-1}\ll\xi/v_F$.
    % Hence for the low-energy theory with a cutoff length scale $a$ satisfying $v_F\beta<a<\xi$,
    % all temporal fluctuations of $\Delta(\bm{r},\bm{s})$ are effectively integrated out.

Under these assumptions, the Hamiltonian $H$ can be regarded as a low-energy effective model
where free electrons experience a spatially disordered pairing potential.
We then solve correlations of $H$ for certain configuration of $\{\Delta(\bm{r},\bm{s})\}$
with perturbative expansion over small $\Delta_0/E_F$,
and take into account the phase fluctuations by considering the following conditions of disorder average,
\begin{subequations}\begin{align}
    \overline{\Delta(\bm{r})} &= \overline{\Delta^\ast(\bm{r})} = 0,\\[3pt]
    \overline{\Delta(\bm{r}) \Delta^\ast(\bm{r^\prime})} &= \Delta_0^2\ g\left(|\bm{r}-\bm{r^\prime}|/\xi\right),
\end{align}\end{subequations}
where the overline represents the average over fluctuating pairing configurations.
$g(x)$ is a general scaleless function that exponentially decays from $g(0)=1$ to 0 as $x\gg1$.
The SC correlation length $\xi$ characterizes the decay of normal-state SC correlations.
In 2D, the SC transition generally falls into the BKT universality class,
and therefore $\xi$ follows a BKT scaling form as
$\xi(T)\sim\exp(bt_r^{-1/2})$,
where $b$ is a non-universal constant and $t_r=(T-T_c)/T_c$ denotes the reduced temperature.

Following the scheme outlined above,
we evaluate the single-particle properties, e.g. the self-energy and electronic spectrum, of $d$-wave SC specifically.
For details of the theoretical calculations, readers may refer to the supplementary material (SM) and Ref.~\cite{wang_phase_2023}.
It turns out that
in the weak-fluctuation regime $T\gtrsim T_c$ with $k_F\xi\gg1$,
the electronic self-energy acquires an intuitive form to the leading order of $\Delta_0$ as,
\begin{equation}\label{eq:approx_self_energy}
    \Sigma(\bm{k},\omega) = \frac{\abs{\Delta_{\bm{k}}}^2}{\omega+\xi_k+2i\gamma_{\bm{k}}}.
\end{equation}
We have defined the $d$-wave gap function $\Delta_{\bm{k}}=\Delta_0\varphi_d(\bm{k})$
and the phase-fluctuation-driven scattering rate $\gamma_{\bm{k}}=v_{\bm{k}}/2\xi$
with $v_{\bm{k}}$ the electronic velocity.
Eq.~\eqref{eq:approx_self_energy} then describes a standard BCS-like self-energy
corrected by a finite quasiparticle lifetime.
Moreover, considering electrons at the Fermi surface, the retarded Green's function 
exhibits complex poles at
\begin{equation}\label{eq:complex_pole_fs}
    E_\pm(\bm{k}) = -i \gamma \pm \sqrt{\abs{\Delta_{\bm{k}}}^2-\gamma^2},
\end{equation}
with the associated pair-scattering rate $\gamma=v_F/2\xi$.

The emergence of pseudogap and Fermi arcs is straightforwardly embedded in the pole structures of Eq.~\eqref{eq:complex_pole_fs}.
Recall that the real components of Eq.~\eqref{eq:complex_pole_fs} denote the quasiparticle energy,
while the imaginary parts characterize the inverse lifetime of quasiparticles.
For temperatures $T$ close to $T_c$ in the normal state,
$\xi(T)$ remains long-ranged such that $\gamma(T)\lesssim\abs{\Delta_{\bm{k}}}$.
In this case, the two Green's function poles manifest at opposite frequencies,
$\text{Re}[E_+]=-\text{Re}[E_-]\neq 0$, as a reminiscence of BCS coherent peaks,
and the pseudogap emerges.
For higher temperatures with relatively short-ranged $\xi(T)$ and $\gamma(T)\gtrsim\abs{\Delta_{\bm{k}}}$,
both of these two quasiparticle modes settle exactly at the Fermi energy,
$\text{Re}[E_\pm]=0$, which corresponds to a metallic spectrum.
As illustrated in Fig.~\ref{fig:fig1}a,
this smooth merging or filling of pseudogap is tracked by either the comparison of $\gamma(T)$ and $\abs{\Delta_{\bm{k}}}$,
or equivalently the competition of BKT correlation length $\xi(T)$ and the BCS coherence length of nodal SC,
$
    \xi_\text{BCS}(\bm{k})=\frac{v_F}{\pi\abs{\Delta_{\bm{k}}}}
$,
which characterizes the spatial extent of Cooper pair in the weak-coupling BCS theory.
A straightforward leading-order calculation in the SM then reveals that
the pseudogap in the electronic single-particle spectrum is not completely closed
until $\frac{\gamma(T)}{\abs{\Delta(\bm{k})}}=\frac{1}{\sqrt{2}}$ or
\begin{equation}\label{eq:crossover}
    \frac{\xi_\text{BCS}(\bm{k})}{\xi(T)} = \frac{\sqrt{2}}{\pi}.
\end{equation}

Furthermore, the above analyses also intuitively explain the presence of Fermi arcs by
taking into account the anisotropy of nodal gap $\abs{\Delta_{\bm{k}}}$,
as shown in Fig.~\ref{fig:fig1}b~and~c.
At a certain temperature in the normal state,
for those momenta near anti-nodes which experience large $\abs{\Delta_{\bm{k}}}$,
the spectrum can be pseudogapped.
Otherwise, for momenta near the nodes, the spectrum remains metallic
and a Fermi surface partially survives at these momenta.
As the temperature increases, the enhanced phase fluctuations rapidly suppress $\xi(T)$
and cause the arcs to grow.
As a result, we have established a universal connection between
the phase fluctuations, pseudogap, and Fermi arcs in 2D nodal superconductors.

It is worth noting that
the analyses above are based on the perturbative expansion over pairing amplitude $\Delta_0$
and the assumption of weak fluctuations $k_F\xi\gg1$.
As inferred from Eq.~\eqref{eq:crossover},
ensuring the validity of our theory in the pseudogap regime further requires that
$\xi\sim\xi_\text{BCS}\gg k_F^{-1}$, which also suggests a small $\Delta_0$.
Consequently, it is reasonable to apply our theory to weak-coupling systems subject to weak phase fluctuations.
Moreover, the Fermi arcs are expected to be generally present for 2D superconductors with nodes,
although we have mostly emphasized the $d$-wave system.
% --------------------------------------------------------------------------

\begin{figure*}[tp]
    \centering\hspace*{-.3cm}
    \includegraphics[width=.9\linewidth]{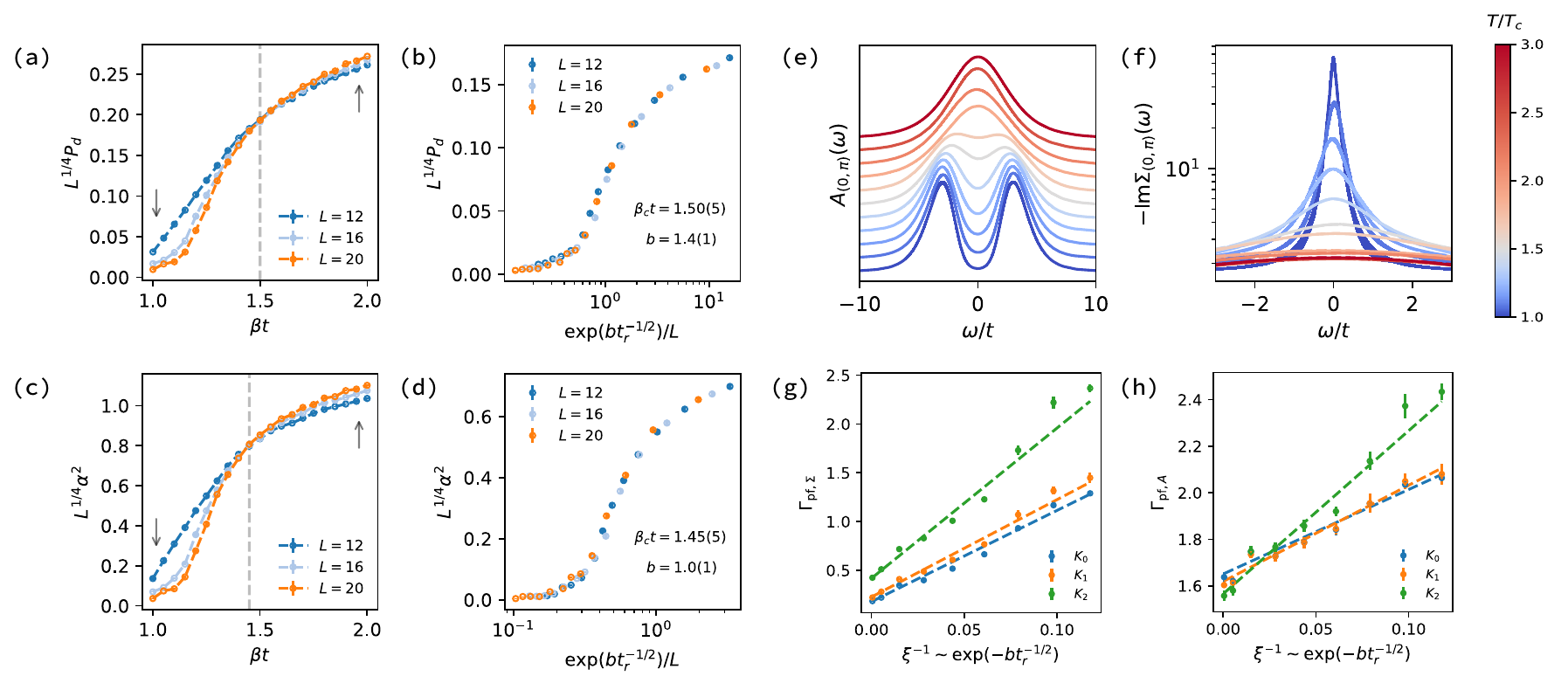}
    \caption{
        (a)(c) Static $d$-wave pairing correlation $P_d$ and $\alpha^2$ for varying inverse temperature $\beta$ and system size $L$.
        We have fixed the critical exponent $\eta(T_c)=1/4$ for the BKT transition.
        (b)(d) Data collapse of $P_d$ and $\alpha^2$.
        (e) Antinodal pseudogap $A_{(0,\pi)}(\omega)$ extracted by combining DQMC and SAC on a $L=20$ lattice.
        (f) Self-energy evaluated from the spectrum in (e). Note that the y-axis is log-scaled for better visualization.
        (g)(h) Pair-scattering rate $\Gamma_\text{pf}$ extracted by
        (g) fitting $\Sigma(\omega)$ to obtain $\Gamma_{\text{pf},\Sigma}$, and
        (h) identifying the half-width of $A(|\omega|\sim\abs{\Delta_k})$ as $\Gamma_{\text{pf},A}$.
        The dashed lines are linear fittings of $\Gamma_\text{pf}$ with respect to $\xi^{-1}$.
        % The DQMC data contributing to the evaluations of electronic spectra and self-energies are obtained on a $L=20$ lattice with $1<T/T_c<1.5$.
        % We have focused on the electronic momenta satisfying $\xi_k=0$ and in the neighborhood of antinodes.
        And the momenta are chosen near the antinodes as $K_n=K_0 + n\Delta k$ with $K_0=(0,\pi)$ and $\Delta k=(2\pi/L,-2\pi/L)$.
    }
    \label{fig:fig2}
\end{figure*}

% {\color{myblue}\it Numerical studies.}
%--------------------------------------------------------------------------
To support our theoretical findings on the phase-fluctuation-driven Fermi arcs,
we investigated a Hubbard-like model on the square lattice,
attributed to Xu and Grover in Ref.~\cite{xu_competing_2021},
with sign-problem-free DQMC simulation.
The model describes standard Hubbard electrons at half-filling,
$
    H_t+H_U = -t \sum_{\left\langle ij\right\rangle,\sigma} ({c}_{i\sigma}^\dag {c}_{j\sigma}+\textrm{h.c.}) + \frac{U}{2} \sum_i ({\rho}_{i\uparrow}+{\rho}_{i\downarrow}-1)^2
$,
coupled to fluctuating $d$-wave $\mathbb{U}(1)$ rotors through
$
    H_V = V \sum_{\langle ij\rangle} [\tau_{ij} \mathrm{e}^{i\theta_{ij}} (c_{i\uparrow}^\dag c_{j\downarrow}^\dag - c_{i\downarrow}^\dag c_{j\uparrow}^\dag) + \textrm{h.c.}]
$.
The dynamics of rotors $\theta_{ij}$ is governed by
$
    H_{XY} = K \sum_{\langle ij\rangle} {\hat{L}_{ij}}^2 - J \sum_{\langle ijk\rangle} \cos\left(\theta_{ij}-\theta_{jk}\right)
$, resembling the quantum rotor model.
As compared to the ground state explored in Ref.~\cite{xu_competing_2021},
we simulate the model at finite temperatures.
And we adopt specific model parameters such that
a nodal $d$-wave superconductor is realized as the ground state, and importantly,
there is no presence of long-range antiferromagnetic order within the temperature range that we considered.
The technical aspects of the model Hamiltonian, model parameters,
and the implementation of DQMC algorithm can be found in the SM.

In Fig.~\ref{fig:fig1}d-g, we show the observation of Fermi arcs
in the electronic single-particle spectrum of the interacting model.
The single-particle spectra are extracted
through the stochastic analytic continuation (SAC) calculations
and the algorithm is briefly described in the SM.
As shown in Fig.~\ref{fig:fig1}g,
in the $d$-wave SC phase, a nodal superconducting gap clearly manifests in the single-particle spectrum,
and the extraction of gap function in Fig.~\ref{fig:fig1}e warrants the $d_{x^2-y^2}$ symmetry of the gap.
As the system enters the normal state with increased temperature,
the Fermi arcs gradually spread to the antinodal regions,
illustrated as red curves in Fig.~\ref{fig:fig1}g.

This evolution of pseudogap and Fermi arcs shall be well attributed to the enhanced phase fluctuations.
To verify this point, we first define the $d$-wave pairing order parameter in terms of both fermionic,
$
    \Delta_d=\frac{1}{2L^2}\sum_{\langle ij\rangle}\tau_{ij}\left(c_{j\downarrow}c_{i\uparrow}-c_{j\uparrow}c_{i\downarrow}\right),
$
and bosonic,
$
    \alpha=\frac{1}{2L^2}\sum_{\langle ij\rangle}e^{i\theta_{ij}},
$
degrees of freedom.
The associated static pairing correlation
$
    P_d=\langle\Delta_d{\Delta_d}^\dag+{\Delta_d}^\dag\Delta_d\rangle
$
and
$
    \alpha^2=\frac{1}{4L^4}\sum_{\langle ij\rangle,\langle kl\rangle} \langle e^{i\theta_{ij}}e^{-i\theta_{kl}}\rangle
$
are then measured and displayed in Fig.~\ref{fig:fig2}a~and~c.
Following the standard procedure of data collapse demonstrated in the SM,
we determine in Fig.~\ref{fig:fig2}b~and~d the inverse critical temperature $\beta_c$
and meanwhile the BKT correlation length $\xi(T)\sim\exp(bt_r^{-1/2})$ up to a prefactor.
$\xi(T)$ is crucial in quantifying the SC fluctuations and serves as the prior input of our theory.
For $P_d$, the data collapse yields $\beta_c t = 1.50(5)$, $b=1.4(1)$,
and for $\alpha^2$, $\beta_c t = 1.45(5)$, $b=1.0(1)$.
It turns out that
the SC transition temperature $T_c=1/\beta_c$ obtained from $P_d$ is consistent with that from $\alpha^2$
within the estimated errorbar.
In contrast, the non-universal factor $b$ varies for $P_d$ and $\alpha^2$.
In the fluctuating regime and at the thermodynamic limit,
physical observables are believed to be governed by a unique BKT correlation length,
leading to consistent scaling exponents.
However, from the perspective of renormalization group and in our finite-size simulations,
the fermionic and bosonic correlation can experience different, effective correlation lengths
before the effective coupling between electrons and rotors flows to the strong-coupling fixed point.
Therefore, this discrepancy in $b$ is expected to asymptotically disappear as
the fixed point is approached by increasing the system size.
Given our primary focus on the electronic side of the model,
we adopt the transition temperature $\beta_c=1.50(5)$ and $b=1.4(1)$ according to the data collapse of $P_d$.

We then validate Eq.~\eqref{eq:approx_self_energy} for the interacting model
by explicitly extracting the electronic self-energy $\Sigma(\omega)$
from the single-particle spectrum $A(\omega)$ obtained through SAC.
This is realized by noting that $A(\omega)=-2\text{Im}[G(\omega)]$
and $\text{Re}[G(\omega)]$ can be computed from $A(\omega)$ based on the Kramers-Kronig relation.
Then the desired self-energy is related to the complex Green's function
through $\Sigma(\omega) = [G^{(0)}]^{-1} - G^{-1}$,
where $G^{(0)}=1/(\omega-\xi_k+i0^+)$ denotes the Green's function of free band electron.
In Fig.~\ref{fig:fig2}e~and~f, we show the spectrum $A(\omega)$ at the antinode obtained from SAC,
and the imaginary part of the evaluated self-energy.
The evolution of pseudogap with respect to $T/T_c$ is clearly illustrated in Fig.~\ref{fig:fig2}e.
Near $T/T_c\gtrsim 1$ and for antinodal electron,
$-\text{Im}[\Sigma(\omega)]$ exhibits a pronounced peak at the Fermi energy $\omega=0$,
subject to a finite broadening caused by scatterings of Cooper pairs.
As $T$ increases,
the self-energy peak in Fig.~\ref{fig:fig2}f gets damped until being completely destroyed by phase fluctuations,
and eventually the pseudogap closes at $T/T_c\sim2$.

To proceed, we fit the self-energies $\Sigma(\omega)$ in Fig.~\ref{fig:fig2}f according to
$
    \Sigma(\omega) = \frac{\widetilde{\Delta}^2}{\omega+2i\Gamma_\text{pf}} - i \Gamma_0,
$
which almost resembles the self-energy in Eq.~\eqref{eq:approx_self_energy} with $\xi_k=0$
for electrons at the Fermi energy.
$\Gamma_\text{pf}$ is interpreted as the pair-scattering rate due to phase fluctuations,
i.e. $\Gamma_\text{pf}(T) \sim v_F\xi^{-1}$.
$\Gamma_0$ can be regarded as a Fermi-liquid-like renormalization,
and $\widetilde\Delta$ denotes the renormalized SC gap.
% It turns out that this form of self-energy is also widely accepted as a working assumption
% in dealing with angle-resolved photoemission data of realistic high-$T_c$ materials~\cite{norman_phenomenology_1998,shi_coherent_2008}.
In practice, we fit the imaginary part of self-energies
for frequencies below the SC gap $\Sigma(|\omega|\lesssim\abs{\Delta_k})$,
as this region dominates the low-energy evolution of the pseudogap.
The fittings are excellent that the fitted curves coincide with the self-energies within errorbars over a broad range of temperatures and momenta.

The fitted pair-scattering rates $\Gamma_{\text{pf},\Sigma}$ were then plotted in Fig.~\ref{fig:fig2}g~and~h,
with respect to $\xi(T)$ obtained previously through data collapse.
Electrons at the Fermi energy but with varying momenta near the antinode are considered.
Ideally it is expected that $\Gamma_{\text{pf},\Sigma}\sim\xi^{-1}$ validating the prediction of our theory.
This linear relation between $\Gamma_\text{pf}$ and $\xi^{-1}$ is well established
over a broad range of temperatures $1<T/T_c<1.5$ in Fig.~\ref{fig:fig2}g,
except for a finite residual of $\Gamma_\text{pf}$ observed at $T_c$ where $\xi^{-1}$ vanishes.
Such residual, denoted as $\Gamma_\text{res}$, is a direct consequence of
the finite half-width of $A(|\omega|\sim\abs{\Delta_k})$ at $T_c$ in Fig.~\ref{fig:fig2}e,
or the finite broadening of self-energy at $T_c$ in Fig.~\ref{fig:fig2}f.
The origin of $\Gamma_\text{res}$ shall be attributed to multiple factors.
Firstly, for a finite-size system, the correlation length is naturally bounded by the system size $L$,
resulting in an effective finite $\xi$ at $T_c$ and hence finite $\Gamma_\text{res}$.
Also, the repulsion among electrons triggers additional scatterings,
and may lead to the additional broadening of quasiparticles at $T_c$.
The inclusion of Hubbard $U$ did not qualitatively alter the Fermi arc formation,
and hence demonstrates the universality of our findings against the interaction effect.
Regardless of the origins of $\Gamma_\text{res}$,
the evolution of $\Gamma_{\text{pf},\Sigma}(T)$ aligns with our theoretical prediction well,
exhibiting a linear dependence on $\xi(T)^{-1}$.

As a supplementary confirmation, 
we note that in the weak-fluctuation regime $\gamma/\abs{\Delta_k}\ll1$,
the half-width of BCS quasiparticle weight $A(|\omega|\sim\abs{\Delta_k})$ in Fig.~\ref{fig:fig2}e
is approximately given by the pair-scattering rate $\gamma$ (as shown in the SM).
It is thus plausible to approximate the pair-scattering rate,
which we denote as $\Gamma_{\text{pf},A}$,
as the half-width of BCS quasiparticle weight near $|\omega|\sim\abs{\Delta_k}$,
although in this case the Fermi-liquid-like correction $\Gamma_0$ is ignored.
Following the similar procedure as for $\Gamma_{\text{pf},\Sigma}$,
we fit the half-width of spectral weights in Fig.~\ref{fig:fig2}e
and again observe in Fig.~\ref{fig:fig2}h the linear relationship between $\Gamma_{\text{pf},A}$ and $\xi^{-1}$.

By combining all these facts, we conclude that
the Fermi arcs can generally arise due to the interplay of normal-state phase fluctuations and nodal quasiparticles in 2D.
The associated BKT correlation length $\xi(T)$ gives rise to the broadening of self-energies $\gamma\sim\xi^{-1}$,
and further its competition with the BCS coherence length $\xi_\text{BCS}$ dominates the evolution of pseudogap and Fermi arcs.
This outlined mechanism is robust against moderate electron repulsion
and quantum correction of pairing fluctuations,
as verified in our Hubbard-like model by DQMC.
Regarding applying our theory to cuprate experiments,
we note that although phase fluctuations alone are insufficient
for interpreting the pseudogap in the underdoped region of hole-doped cuprates,
which might be more relevant to competing orders,
a ubiquitous fluctuating SC region and the phase-fluctuation-driven pseudogap
are experimentally identified~\cite{sobota_angle-resolved_2021}
over the entire superconducting dome on the hole-doped side,
where our theory is generally applicable.
As for the electron-doped side, signatures of fluctuating SC are rarely reported.
We attribute this to the loss of effective two-dimensionality,
indicated by significantly larger superfluid stiffness observed in electron-doped compounds.

% It is however notable that
% beyond the single-particle property explored primarily in this study,
% it is equally compelling to examine the two-particle properties,
% such as the response to external magnetic probes.
% Moreover, the framework here can be readily extended to address problems
% concerning static disorders and other fluctuating orders, e.g. spin fluctuations, at finite temperatures.
% All these directions will guide our future works.
%--------------------------------------------------------------------------
\vspace{.5cm}

\newcommand{\nocontentsline}[3]{}
\let\origcontentsline\addcontentsline
\newcommand\stoptoc{\let\addcontentsline\nocontentsline}
\newcommand\resumetoc{\let\addcontentsline\origcontentsline}

\stoptoc

%--------------------------------------------------------------------------
{\color{myblue}\noindent \textbf{Conflict of interest}}

\vspace{.3cm}
The authors declare that they have no conflict of interest.
\vspace{.3cm}
%--------------------------------------------------------------------------

%--------------------------------------------------------------------------
{\color{myblue}\noindent \textbf{Acknowledgements}}

\vspace{.3cm}
This work is supported by the National Key R\&D Program of China (Grant Nos. 2022YFA1402702, 2021YFA1401400 and 2022YFA1403402), the National Natural Science Foundation of China (Grant Nos. 12447103, 12274289, 12174068 and 12374144), the Innovation Program for Quantum Science and Technology (Grant No. 2021ZD0301902) and the Science and Technology Commission of Shanghai Municipality (Grant Nos. 24LZ1400100 and 23JC1400600).
Xiao Yan Xu acknowledges the support of Yangyang Development Fund, Shanghai Jiao Tong University 2030 Initiative, and startup funds from SJTU.
Yang Qi acknowledges the support of Shuguang Program of Shanghai Education Development Foundation and Shanghai Municipal Education Commission.
The authors also appreciate \href{https://cloud.paratera.com}{Beijing PARATERA Tech CO., Ltd.} for providing the HPC resources, which have contributed to the computational results presented in this work.
\vspace{.3cm}
%--------------------------------------------------------------------------

%--------------------------------------------------------------------------
{\color{myblue}\noindent \textbf{Author contributions}}

\vspace{.3cm}
Xiao Yan Xu and Yang Qi initiated the project.
Xu-Cheng Wang performed the theoretical and numerical calculations.
All authors contributed to the data analysis and were engaged in the writing of the manuscript.
\vspace{.3cm}
%--------------------------------------------------------------------------

\bibliographystyle{apsrev4-2}
\bibliography{ref.bib}

\resumetoc

\newpage
\clearpage
\onecolumngrid

\begin{center}
%--------------------------------------------------------------------------
\textbf{
Supplementary Material for
``The interplay of phase fluctuations and nodal quasiparticles:
ubiquitous Fermi arcs in two-dimensional $\textit{d}$-wave superconductors"
}
%--------------------------------------------------------------------------
\end{center}

\setcounter{equation}{0}
\setcounter{figure}{0}
\setcounter{table}{0}
\setcounter{page}{1}

\makeatletter

\renewcommand{\thetable}{S\arabic{table}}
\renewcommand{\theequation}{S\arabic{equation}}
\renewcommand{\thefigure}{S\arabic{figure}}
\renewcommand{\bibnumfmt}[1]{[S#1]}
\renewcommand{\citenumfont}[1]{S#1}
\setcounter{secnumdepth}{3}

\tableofcontents

\section{Perturbative Evaluation of the Electronic Self-energy}
\begin{figure}[bp]
    \centering
    \includegraphics[width=.3\textwidth]{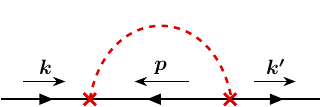}
    \caption{
        \label{fig:figs1}
        Feynman diagram for the second-order self-energy correction.
        The arrowed solid lines denote the bare propagators of band electrons,
        while the dashed red line represents the disorder average over the fluctuating pairing correlations.
    }
\end{figure}

%--------------------------------------------------------------------------
In this section, we derive the electronic self-energy of the phase-fluctuating superconductor
by combining perturbation theory and disorder-average technique~\cite{mahan_many_2000}.
Due to the disorder-average condition $\overline{\Delta(\bm{r})}=\overline{\Delta^\ast(\bm{r})}=0$,
the first-order correction simply vanishes.
Hence, we focus on the second-order correction,
which follows straightforwardly from the Feynman diagram in Fig.~\ref{fig:figs1} and is written down as
\begin{equation}\label{eq:self_energy}
    \Sigma(\bm{k},\bm{k^\prime},\omega) = \frac{1}{V} \int\frac{\mathrm{d}^2\bm{p}}{\left(2\pi\right)^2} \frac{\Delta_{\bm{kp}}\Delta_{\bm{pk^\prime}}^\ast}{\omega+\xi_p+i0^+}.
\end{equation}
The spin index will be omitted throughout the calculation since the self-energy is trivial in the spin subspace.

The self-energy Eq.~\eqref{eq:self_energy} depends on
both the incoming momentum $\bm{k}$ and the outcoming one $\bm{k^\prime}$
because the translational invariance is explicitly broken
for certain fixed pairing configurations $\{\Delta(\bm{r},\bm{s})\}$.
The Fourier-transformed pairing order parameter $\Delta_{\bm{kp}}$ is defined as
\begin{equation}\begin{aligned}
    \Delta_{\bm{p_1p_2}}
    &= \int\mathrm{d}^2\bm{r}\mathrm{d}^2\bm{s}\ \Delta(\bm{r},\bm{s})\ e^{-i\bm{p_1}(\bm{r}+\bm{s}/2)-i\bm{p_2}(\bm{r}-\bm{s}/2)}\\
    &= \int\mathrm{d}^2\bm{r}\ \Delta(\bm{r})\ e^{-i(\bm{p_1}+\bm{p_2})\bm{r}}\ \varphi\left(\frac{\bm{p_1}-\bm{p_2}}{2}\right),
\end{aligned}\end{equation}
where in the second step we have applied the relation
$\int\mathrm{d}^2\bm{s}\ \mathrm{e}^{-i\bm{p}\bm{s}}\Delta(\bm{r},\bm{s}) = \Delta(\bm{r})\varphi(\bm{p})$.
Hereafter, we will assume a $d$-wave form factor $\varphi(\bm{p})=(p_x^2-p_y^2)/p^2$,
but note that a similar analysis can be conducted to deal with generic pairing symmetries.
In addition, for the convenience of analytic calculation and without losing generality,
we adopt a Gaussian decay function
$g(|\bm{r}-\bm{r^\prime}|/\xi)=e^{-\abs{\bm{r}-\bm{r^\prime}}^2/2\xi^2}$
for the disorder-averaged two-point pairing correlation,
and express it in the momentum space as
\begin{equation}\begin{aligned}
    \overline{\Delta_{\bm{p_1}\bm{p_2}}\Delta_{\bm{p_3}\bm{p_4}}^\ast} &= \int\mathrm{d}^2\bm{r_1}\mathrm{d}^2\bm{r_2}\ \overline{\Delta(\bm{r_1})\Delta(\bm{r_2})^\ast}\ e^{-i(\bm{p_1}+\bm{p_2})\bm{r_1}+i(\bm{p_3}+\bm{p_4})\bm{r_2}}\ \varphi\left(\frac{\bm{p_1}-\bm{p_2}}{2}\right)\varphi^*\left(\frac{\bm{p_3}-\bm{p_4}}{2}\right)\\[5pt]
    &= 2\pi\xi^2\Delta_0^2\ V\delta^{(2)}\left(\bm{p_1}+\bm{p_2}-\bm{p_3}-\bm{p_4}\right) e^{-\left(\bm{p_1}+\bm{p_2}\right)^2\xi^2/2}\ \varphi\left(\frac{\bm{p_1}-\bm{p_2}}{2}\right)\varphi^*\left(\frac{\bm{p_3}-\bm{p_4}}{2}\right).
\end{aligned}\end{equation}
Substituting this result into Eq.~\eqref{eq:self_energy},
we obtain the disorder-averaged self-energy
$\overline{\Sigma(\bm{k},\bm{k^\prime},\omega)}=\Sigma(\bm{k},\omega)\delta^{(2)}(\bm{k}-\bm{k^\prime})$ with
\begin{equation}\label{eq:self_energy_s4}
    \Sigma(\bm{k},\omega) = 2\pi\xi^2 \Delta_0^2 \int\frac{\mathrm{d}^2\bm{p}}{(2\pi)^2} \frac{e^{-(\bm{k}+\bm{p})^2\xi^2/2}}{\omega+\xi_p+i0^+}\ \abs{\varphi\left(\frac{\bm{k}-\bm{p}}{2}\right)}^2.
\end{equation}
As expected, the translational invariance is recovered after averaging over the fluctuating pairing potential.
From Eq.~\eqref{eq:self_energy_s4}, one can already realize that
the self-energy acquires a finite imaginary component due to the scattering mediated by phase fluctuations,
contributing to the finite life-time of Cooper pairs.
This can be brought out explicitly by further considering the weakly fluctuating regime with $k_F\xi\gg1$,
or equivalently temperature range near $T_c$ in the normal state,
and examining the imaginary part of Eq.~\eqref{eq:self_energy_s4},
\begin{equation}\begin{aligned}\label{eq:self_energy_s5}
    \text{Im}\Sigma(\bm{k},\omega) &\approx 2\pi\xi^2 \abs{\Delta_{\bm{k}}}^2 \int\frac{\mathrm{d}^2\bm{p}}{(2\pi)^2}\ e^{-(\bm{k}+\bm{p})^2\xi^2/2} \left[-\pi\delta(\omega+\xi_p)\right]\\
    &= -\pi m\xi^2 \abs{\Delta_{\bm{k}}}^2\ e^{-\xi^2k^2+m\xi^2(\omega+\xi_k)} I_0\left(\xi^2k^2\sqrt{1-\frac{2m}{k^2}(\omega+\xi_k)}\right), \quad \text{with} \quad \omega<\mu\\
    &\approx -\pi \abs{\Delta_{\bm{k}}}^2\ \frac{m\xi}{(2\pi)^{1/2}k}\ e^{-\frac{m^2\xi^2}{2k^2}(\omega+\xi_k)^2}.
\end{aligned}\end{equation}
For finally arriving at Eq.~\eqref{eq:self_energy_s5},
we first observe that the integrand in Eq.~\eqref{eq:self_energy_s4} takes significant value
only for a small portion of momenta around $\bm{p}=-\bm{k}$ with a soft cutoff $\xi^{-1}$.
This allows us to substitute $\varphi[(\bm{k}-\bm{p})/2]$ in Eq.~\eqref{eq:self_energy_s4} with $\varphi[\bm{k}]$,
and further absorb it into $\Delta_{\bm{k}}$ in the first step of Eq.~\eqref{eq:self_energy_s5}.
Also, we have used the identity $\text{Im}[1/(\omega+\xi_p+i0^+)]=-\pi\delta(\omega+\xi_p)$.
In the second step, the integral over $\bm{p}$ is solved exactly by assuming the dispersion $\xi_k=k^2/2m-\mu$,
and $I_0$ is the zero-order modified Bessel function of the first kind.
The asymptotic form of $I_0$ in the $k_F\xi\gg1$ limit is adopted in the last step,
and it is finally found that the imaginary part of the self-energy obeys a Gaussian distribution
peaking at $\omega=-\xi_k$ with the standard deviation $\sigma=2\gamma_{\bm{k}}\overset{!}{=} k/m\xi$.

\begin{figure}[htbp]
    \centering
    \includegraphics[width=.9\textwidth]{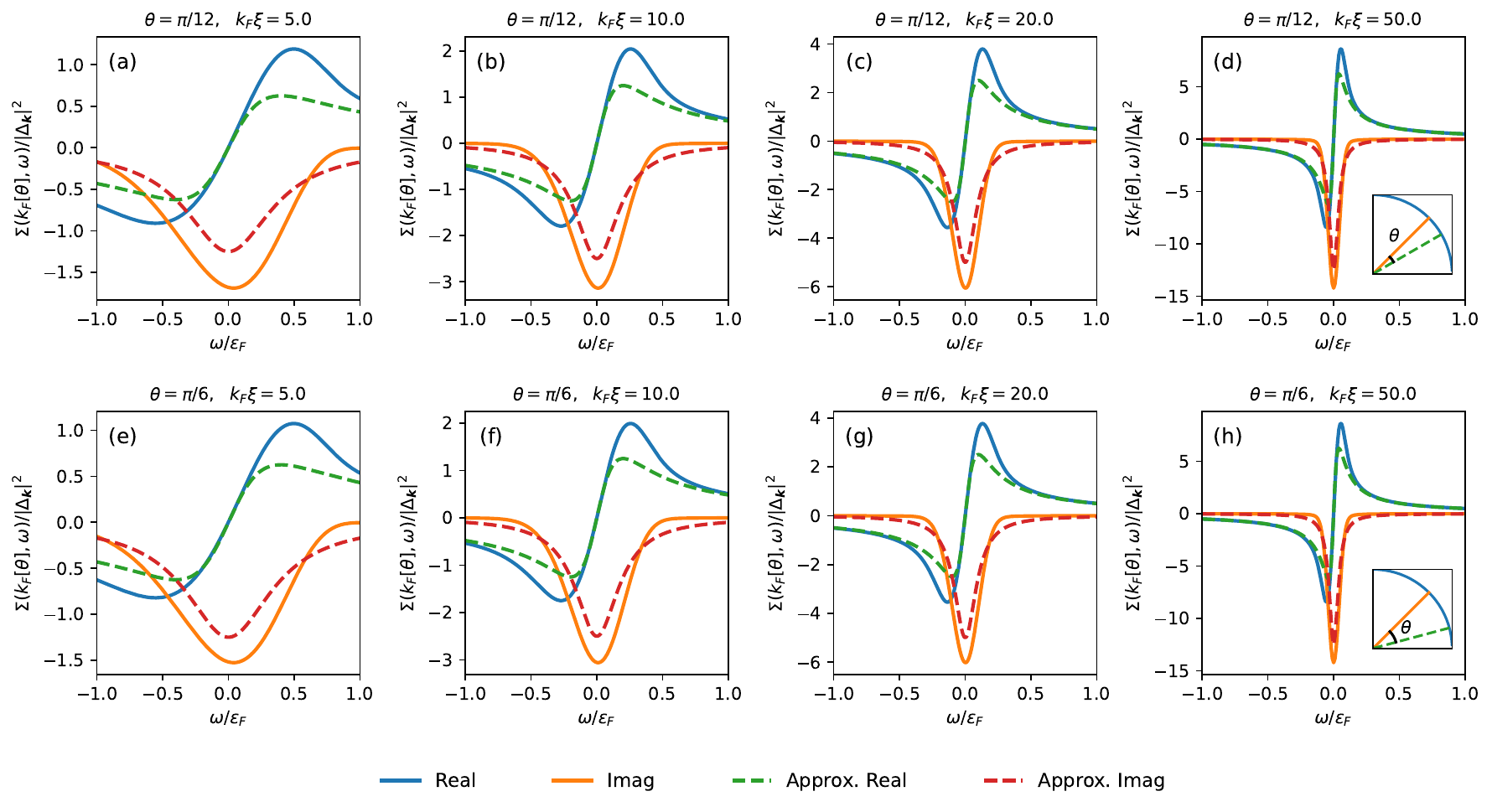}
    \caption{
        Benchmark of self-energies in Eq.~\eqref{eq:self_energy_s4} and Eq.~\eqref{eq:self_energy_s6}
        by varying both the correlation length $\xi$ and the Fermi momentum $\bm{k}_F[\theta]$.
        $\theta$ denotes the angle to the $d$-wave node, as illustrated in the mini charts of the rightmost figures.
        The exact self-energies, Eq.~\eqref{eq:self_energy_s4}, are plotted in solid lines while the approximated ones, Eq.~\eqref{eq:self_energy_s6}, are in dashed lines.
        It is found that Eq.~\eqref{eq:self_energy_s6} produces similar real and imaginary components as those of Eq.~\eqref{eq:self_energy_s4}, especially when $k_F\xi$ grows up.
        We have also tested Eq.~\eqref{eq:self_energy_s6} for momenta away from the Fermi surface,
        which only causes a frequency shift as compared to the results above,
        and hence we do not show them separately.
    }
    \label{fig:figs2}
\end{figure}

Eq.~\eqref{eq:self_energy_s5} then 
motivates us to intuitively rewrite the self-energy Eq.~\eqref{eq:self_energy_s4}
in the weakly fluctuating regime as
\begin{equation}\label{eq:self_energy_s6}
    \Sigma(\bm{k},\omega) = \frac{\abs{\Delta_{\bm{k}}}^2}{\omega+\xi_k+2i\gamma_{\bm{k}}},
\end{equation}
which is exactly what we have announced in the main text.
The imaginary part of Eq.~\eqref{eq:self_energy_s6} now obeys a Lorentz distribution
as compared with the Gaussian one in Eq.~\eqref{eq:self_energy_s5}.
For temperature $T$ close to $T_c$ in the normal state,
it is plausible that Eq.~\eqref{eq:self_energy_s4} and Eq.~\eqref{eq:self_energy_s6} shall behave similarly.
To see this, both the real and imaginary parts of Eq.~\eqref{eq:self_energy_s4} and Eq.~\eqref{eq:self_energy_s6} are benchmarked numerically in Fig.~\ref{fig:figs2},
where good agreements are achieved for $k_F\xi\gg1$.
We argue that for any specific form of two-point pairing correlation $g(r/\xi)$,
the detailed high-energy information of the correlation decay can hardly be probed in the vicinity of $T_c$, and hence turns out to be irrelevant.
Under this assumption, a similar analysis as we outline above can be applied for generic pairing correlations,
and it is generally expected that the effects of weak phase fluctuations
shall be well captured by the self-energy in Eq.~\eqref{eq:self_energy_s6}.

%--------------------------------------------------------------------------

\section{Evolution of Electronic Single-particle Spectrum and the Pseudogap}
\begin{figure}[htbp]
    \centering\hspace*{-.5cm}
    \includegraphics[width=0.55\textwidth]{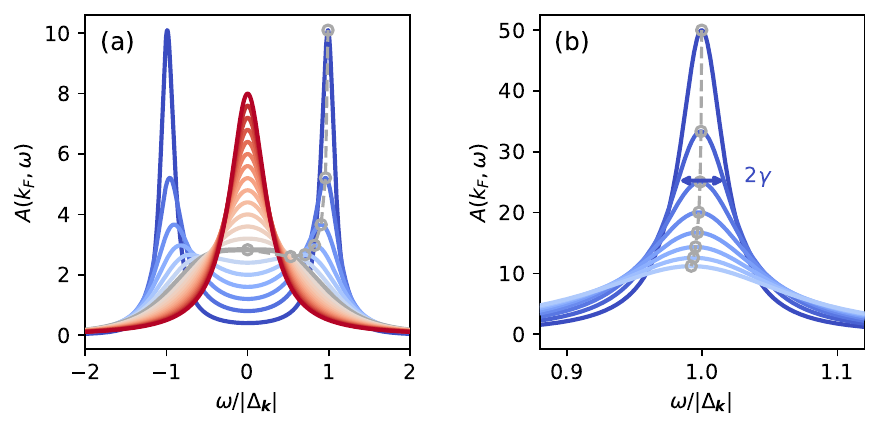}
    \caption{
        Single-particle spectral function $A(\bm{k}_F,\omega)$ at the Fermi momentum for varying $\gamma/\abs{\Delta_{\bm{k}}}$.
        We outline the spectral peaks at $\omega=\Delta_\text{pg}=\sqrt{\abs{\Delta_{\bm{k}}}^2-2\gamma^2}$ using the dashed line with circles.
        (a) The crossover from the BCS physics $(\gamma\lesssim\abs{\Delta_{\bm{k}}})$ to the Fermi liquid $(\gamma\gtrsim\abs{\Delta_{\bm{k}}})$
            with $\gamma/\abs{\Delta_{\bm{k}}}$ ranging from $10^{-1}$ to $2$.
            The pseudogap develops in the intermediate regime of this crossover,
            and the energy gap is fully closed when $\gamma/\abs{\Delta_{\bm{k}}}=1/\sqrt{2}$,
            noted as the grey solid curve.
        (b) Broadening of the BCS quasiparticle peak, e.g. $A(\omega\sim\abs{\Delta_{\bm{k}}})$,
            in the weakly fluctuating regime for $\gamma/\abs{\Delta_{\bm{k}}}$ ranging from $2\times10^{-2}$ to $10^{-1}$.
            The two-sided arrow implies that the spectral peak acquires a half-width proportional to $\gamma$.
    }
    \label{fig:figs3}
\end{figure}

%--------------------------------------------------------------------------
Following directly from the self-energy in Eq.~\eqref{eq:self_energy_s6},
the retarded Green's function is written down as
\begin{equation}\label{eq:greens_func}
    G(\bm{k},\omega) = \left[\omega-\xi_k-\Sigma(\bm{k},\omega)\right]^{-1} = \frac{\omega+\xi_k+2i\gamma_{\bm{k}}}{\omega^2-E_k^2+2i\gamma_{\bm{k}}\left(\omega-\xi_k\right)},
\end{equation}
where we define the $d$-wave BCS dispersion $E_k=\sqrt{\xi_k^2+\abs{\Delta_{\bm{k}}}^2}$.
The quasiparticle dispersion in the presence of phase fluctuations
is embedded in the complex poles of Green's function,
which are readily read from Eq.~\eqref{eq:greens_func},
\begin{equation}\label{eq:complex_pole}
    E_\pm(\bm{k}) = -i\gamma_{\bm{k}} \pm \sqrt{\left(\xi_k+i\gamma_{\bm{k}}\right)^2+\abs{\Delta_{\bm{k}}}^2}.
\end{equation}
For electrons with Fermi momentum $\bm{k}_F$, i.e. $\xi_{\bm{k}_F}=0$,
Eq.~\eqref{eq:complex_pole} then recovers the result in the main text.
Moreover, the single-particle spectral function $A(\bm{k},\omega)$ is related to the imaginary component of $G(\bm{k},\omega)$ as
\begin{equation}
    A(\bm{k},\omega) = -2\text{Im}[G(\bm{k},\omega)] = \frac{4\abs{\Delta_{\bm{k}}}^2\gamma_{\bm{k}}}{\left(\omega^2-E_k^2\right)^2+4\gamma_{\bm{k}}^2\left(\omega-\xi_k\right)^2},
\end{equation}
and for Fermi momentum,
\begin{equation}\label{eq:spec_func_fs}
    A(\bm{k}_F,\omega) = \frac{4\abs{\Delta_{\bm{k}}}^2\gamma}{\left(\omega^2-\abs{\Delta_{\bm{k}}}^2\right)^2+4\gamma^2\omega^2},
\end{equation}
where the pair-scattering rate $\gamma=v_F/2\xi$.
By examining the denominator of Eq.~\eqref{eq:spec_func_fs},
it is easy to find that $A(\bm{k}_F,\omega)$ reaches its maximum when
\begin{subequations}\begin{align}
        \omega = 0, &\qquad \gamma/\abs{\Delta_{\bm{k}}}\geqslant 1/\sqrt{2};\\[5pt]
        \omega = \pm\sqrt{\abs{\Delta_{\bm{k}}}^2-2\gamma^2}, &\qquad \gamma/\abs{\Delta_{\bm{k}}}<1/\sqrt{2}.
\end{align}\end{subequations}
According to these observations, we define the energy scale of pseudogap
$\Delta_\text{pg}(\bm{k})=\sqrt{\abs{\Delta_{\bm{k}}}^2-2\gamma^2}$.
It is then realized that above $T_c$ in the normal state,
the spectral gap remains open unless $\gamma/\Delta_{\bm{k}}\geqslant 1/\sqrt{2}$.
The smooth crossover from the BCS physics when $T\gtrsim T_c$
to the high-temperature Fermi liquid when $T\gg T_c$,
i.e. the evolution of pseudogap, is depicted in Fig.~\ref{fig:figs3}(a).
In the weakly fluctuating regime with $\gamma\ll\abs{\Delta_{\bm{k}}}$,
we infer from Eq.~\eqref{eq:spec_func_fs} that
$
    A_{\pm}(\bm{k}_F,\omega)\approx\gamma/\left[\left(\omega\pm\abs{\Delta_{\bm{k}}}\right)^2+\gamma^2\right],
$
where the half-width broadening of BCS quasiparticle weights
$
    A\left(|\omega|\sim\abs{\Delta_{\bm{k}}}\right)
$
are simply characterized by $\gamma$, as shown in Fig.~\ref{fig:figs3}(b).

%--------------------------------------------------------------------------

\section{Implementation of Determinant Quantum Monte Carlo Algorithm}
%--------------------------------------------------------------------------
\subsection{Model Hamiltonian}
As proposed in Ref.~\cite{xu_competing_2021}, the model Hamiltonian is described by
\begin{subequations}\label{eq:hubbard-rotor-model}\begin{align}
    H_t &= -t \sum_{\left\langle ij\right\rangle,\sigma} \left(c_{i\sigma}^\dag c_{j\sigma}+\textrm{h.c.}\right),\\
    H_U &= \frac{U}{2} \sum_i \left(\rho_{i\uparrow}+\rho_{i\downarrow}-1\right)^2,\\
    H_V &= V \sum_{\langle ij\rangle} \left[\tau_{ij} \mathrm{e}^{i\theta_{ij}} \left(c_{i\uparrow}^\dag c_{j\downarrow}^\dag - c_{i\downarrow}^\dag c_{j\uparrow}^\dag\right) + \textrm{h.c.}\right],\\[4pt]
    H_{XY} &= K \sum_{\langle ij\rangle} {\hat{L}_{ij}}^2 - J \sum_{\langle ijk\rangle} \cos\left(\theta_{ij}-\theta_{jk}\right).
\end{align}\end{subequations}
The model involves band electrons $c_{i,\sigma}$($c^\dag_{i,\sigma}$),
which live on the vertices $\{i\}$ of the square lattice,
and the fluctuating $\mathbb{U}(1)$ rotors $e^{i\theta_{ij}}$ living on the nearest-neighbor bonds $\{ij\}$.
$H_t+H_U$ denotes the standard Hubbard model at half-filling with nearest-neighbor hoppings,
and $\rho_{i\sigma}=c_{i\sigma}^\dag c_{i\sigma}$.
$H_V$ describes the coupling between electrons and rotors in the $d$-wave pairing channel,
where $\tau_{ij}=1$ for bonds $\left\langle ij\right\rangle$ along the $x$-axis and $-1$ for bonds along the $y$-axis.
$H_{XY}$ then governs the dynamics of rotors
and a self-interaction term is included resembling the one in the quantum rotor model~\cite{sachdev_quantum_2011}.
$\hat{L}_{ij}$ denotes the canonical momentum of rotor,
and $\sum_{\langle ijk\rangle}$ sums over pairs of rotors which share the same lattice site $j$.

Note that this model preserves all symmetries present in the half-filled Hubbard model,
including the spin-rotation $\mathbb{SU}(2)$, particle-hole and the $\mathbb{U}(1)$ charge symmetry.
In the absence of Hubbard repulsion, it simply describes, in a minimal way,
the free electrons coupled to BKT superconducting fluctuations,
which is expected as the universal low-energy effective description
to any $d$-wave superconductor that exhibits BKT physics.
Adding the Hubbard term $H_U$ further mimics realistic models by incorporating Mott physics.
In the present work, we include moderate Hubbard repulsion to demonstrate the generality of our findings
on phase-fluctuation-driven Fermi arcs in the presence of interaction effect.
In this sense, although the model is restricted to half-filling,
it still maintains relevance to basic experimental setups
and preserves exact solvability through avoiding the sign problem.
(Introducing any finite doping would inevitably result in the sign problem.)

In Ref.~\cite{xu_competing_2021}, a complete ground-state phase diagram of this model has been established:
it hosts three distinct phases, including a nodal $d$-wave phase,
which is the focus of our interest here,
an antiferromagnet,
and an intervening phase where AFM and nodeless dSC coexist.
In contrast, we explore in this work the finite-temperature properties, especially the BKT transition and normal-state electronic spectrum.
Unless otherwise specified,
we set the imaginary-time spacing $\Delta\tau$ equal to $0.025/t$
for maintaining a controllable systematic error, and work with the lattice size $L$ ranging from 12 to 20.
For all our presented results, we have set $t=1$, $U/t=4$, $V/t=1$, and $K/t=J/t=1/2$,
which is shown to realize a dSC ground state and host a BKT transition to the normal state.
In Sec.~\ref{sec:no-afm}, we further prove that there are no signals of AFM order under this set of parameters.

\subsection{Basic Formalism of Determinant Quantum Monte Carlo}
In this section, we discuss the determinant Quantum Monte Carlo (DQMC) simulation of
the model described by Eqs.~\eqref{eq:hubbard-rotor-model}.
It is insightful to first switch to the particle-hole channel,
and reformulate Eqs.~\eqref{eq:hubbard-rotor-model} in a transformed basis
$
\left(\tilde{c}^\dag_{i\uparrow},\tilde{c}^\dag_{i\downarrow}\right) \overset{!}{=} \left(c^\dag_{i\uparrow},(-)^ic_{i\downarrow}\right)
$
as
\begin{subequations}\begin{align}
    H_t &= -t \sum_{\left\langle ij\right\rangle,\sigma} \left(\tilde{c}_{i\sigma}^\dag \tilde{c}_{j\sigma}+\textrm{h.c.}\right),\\
    H_U &= -\frac{U}{2} \sum_i \left(\tilde{\rho}_{i\uparrow}+\tilde{\rho}_{i\downarrow}-1\right)^2,\\
    H_V &= -V \sum_{\langle ij\rangle} \left[\tau_{ij} \mathrm{e}^{i\theta_{ij}} (-)^i \left(\tilde{c}_{i\uparrow}^\dag \tilde{c}_{j\downarrow} - \tilde{c}_{j\uparrow}^\dag \tilde{c}_{i\downarrow}\right) + \textrm{h.c.}\right],\\[4pt]
    H_{XY} &= K \sum_{\langle ij\rangle} {\hat{L}_{ij}}^2 - J \sum_{\langle ijk\rangle} \cos\left(\theta_{ij}-\theta_{jk}\right),
\end{align}\end{subequations}
where $\tilde{\rho}_{i\sigma}=\tilde{c}^\dag_{i\sigma}\tilde{c}_{i\sigma}$ denotes the density operator in the particle-hole basis.
Within the framework of DQMC and given the partition function $\mathcal{Z}=\text{Tr}\left(e^{-\beta H}\right)$,
one first divides the imaginary-time evolution $\beta$ into $N_\tau$ slices with imaginary-time spacing $\Delta\tau=\beta/N_\tau$.
To further isolate the quartic interaction terms from the quadratic ones,
we decompose the exponent of $H$ at a certain time slice $\tau$ into more components, where different components basically do not commute.
This is fulfilled by applying Trotter decomposition at the cost of an accompanying systematic error of order $O[(\Delta\tau)^2]$, as in Eq.~\eqref{eq:partition_function},
\begin{equation}\label{eq:partition_function}
    \mathcal{Z} = \text{Tr}\left[\left(e^{-\Delta\tau H}\right)^{N_\tau}\right]
    \approx \text{Tr}\left[\left(e^{-\Delta\tau H_{XY}}e^{-\Delta\tau H_U}e^{-\Delta\tau H_V}e^{-\Delta\tau H_t}\right)^{N_\tau}\right].
\end{equation}
To analytically integrate out the fermionic degrees of freedom, one needs to transform the two-body interaction term $H_U$ into fermion bilinears under the Hubbard-Stratonovich (HS) transformation.
The resulting fermion bilinears are coupled to certain auxiliary bosonic fields $\{s_{i,\tau}\}$, which take values in $\{\pm{1},\pm{2}\}$~\cite{assaad_phase_2005},
\begin{equation}\label{eq:hs_decomposition}
    e^{-\Delta\tau H_U(\tau)} = \prod_{i} e^{\Delta\tau\frac{U}{2}\left[\tilde{\rho}_{i\uparrow}(\tau)+\tilde{\rho}_{i\downarrow}(\tau)-1\right]^2}
    \approx \frac{1}{4^N} \prod_{i}\sum_{s_{i,\tau}} \gamma(s_{i,\tau}) e^{\alpha\eta(s_{i,\tau})\left[\tilde{\rho}_{i\uparrow}(\tau)+\tilde{\rho}_{i\downarrow}(\tau)-1\right]},
\end{equation}
where $N=L\times L$ counts the total number of lattice sites, $\alpha=\sqrt{\Delta\tau\frac{U}{2}}$, $\gamma(\pm{1})=1+\sqrt{6}/3$, $\gamma(\pm{2})=1-\sqrt{6}/3$, $\eta(\pm{1})=\pm\sqrt{6-2\sqrt{6}}$, and $\eta(\pm{2})=\pm\sqrt{6+2\sqrt{6}}$.
This particular scheme of HS transformation preserves the spin $\mathbb{SU}(2)$ symmetry
and introduces a systematic error of order $O[(\Delta\tau)^3]$ when physical observables are measured,
which is, however, neglectable compared with the Trotter error.

To deal with the bosonic rotor fields $\hat{\theta}_{ij}$ present in $H_{XY}$ in the path integral formalism,
it is convenient to adopt the bosonic coherent state representation $\left\{\ket{\theta}\right\}$
satisfying $\hat{\theta}_{ij}\ket{\theta}=\theta_{ij}\ket{\theta}$ for all bonds $\langle ij\rangle$.
Then our remaining task is to estimate the matrix elements
$\bra{\theta^\prime}e^{-\Delta\tau K\sum_{\langle ij\rangle}{\hat{L}_{ij}}^2}\ket{\theta}$.
This can be achieved by inserting the completeness relation of angular momentum eigenstates $\{\ket{L}\}$ and applying the Poisson summation formula~\cite{jiang_monte_2022}.
As a result, under the Villain approximation, we will arrive at
\begin{equation}\label{eq:rotor_dynamics}
    \bra{\theta^\prime}e^{-\Delta\tau K\sum_{\langle ij\rangle}{\hat{L}_{ij}}^2}\ket{\theta} \sim e^{\frac{1}{2\Delta\tau K}\sum_{\langle ij\rangle}\cos{\left(\theta_{ij}-\theta_{ij}^\prime\right)}}.
\end{equation}
After all these preparations, we are now ready to evaluate the partition function in Eq.~\eqref{eq:partition_function}.
First of all, we trace out the bosonic fields using Eq.~\eqref{eq:rotor_dynamics} and the properties of coherent states,
\begin{equation}\begin{aligned}
    \mathcal{Z} &= \text{Tr}_F \sum_{\{\theta_{ij,\tau}\}} \bra{\theta_{\tau=1}}e^{-\Delta\tau H}\ket{\theta_{\tau=N_\tau}}
                   \bra{\theta_{\tau=N_\tau}}e^{-\Delta\tau H}\ket{\theta_{\tau=N_\tau-1}} \cdots
                   \bra{\theta_{\tau=2}}e^{-\Delta\tau H}\ket{\theta_{\tau=1}} \\
                &= \sum_{\{\theta_{ij,\tau}\}} e^{\frac{1}{2\Delta\tau K}\sum_{\left\langle ij\right\rangle,\tau}\cos\left(\theta_{ij,\tau+1}-\theta_{ij,\tau}\right) + \Delta\tau J\sum_{\langle ijk\rangle,\tau}\cos\left(\theta_{ij,\tau}-\theta_{jk,\tau}\right)}
                   \ \text{Tr}_F \left[\prod_{\tau}\left(e^{-\Delta\tau H_U}e^{-\Delta\tau H_V}e^{-\Delta\tau H_t}\right)\right],
\end{aligned}\end{equation}
where we have mapped the quantum rotor term $H_{XY}$ into a (2+1)D anisotropic XY model at the path integral level.
Furthermore, we trace out the fermion bilinears using Eq.~\eqref{eq:hs_decomposition},
\begin{equation}\begin{aligned}\label{eq:trace_fermion}
    \text{Tr}_F \left[\prod_{\tau}\left(e^{-\Delta\tau H_U}e^{-\Delta\tau H_V}e^{-\Delta\tau H_t}\right)\right] &= 
    \sum_{\{s_{i,\tau}\}} \left(\prod_{i,\tau}\gamma(s_{i,\tau})e^{-\alpha\eta(s_{i,\tau})}\right) \text{Tr}_F \left[\prod_{\tau}\left(e^{\mathbf{\tilde{c}}^\dag \mathbf{K}_U \mathbf{\tilde{c}}}e^{\mathbf{\tilde{c}}^\dag \mathbf{K}_V \mathbf{\tilde{c}}}e^{\mathbf{\tilde{c}}^\dag \mathbf{K}_t \mathbf{\tilde{c}}}\right)\right]\\
    &= \sum_{\{s_{i,\tau}\}} \left(\prod_{i,\tau}\gamma(s_{i,\tau})e^{-\alpha\eta(s_{i,\tau})}\right) \text{det}\left(\mathbf{1}+\prod_{\tau}\mathbf{B}_\tau\right),
\end{aligned}\end{equation}
with the constant prefactor omitted.
In Eq.~\eqref{eq:trace_fermion}, $\mathbf{\tilde{c}}=(\tilde{c}_{1,\uparrow},\cdots,\tilde{c}_{N,\uparrow},\tilde{c}_{1,\downarrow},\cdots,\tilde{c}_{N,\downarrow})^T$ denotes our fermionic basis,
and the three matrices $\mathbf{K}_t$, $\mathbf{K}_V$, $\mathbf{K}_U$ satisfy that $\mathbf{\tilde{c}}^\dag \mathbf{K}_t \mathbf{\tilde{c}}=-\Delta\tau H_t$, $\mathbf{\tilde{c}}^\dag \mathbf{K}_V \mathbf{\tilde{c}}=-\Delta\tau H_V$, and $\mathbf{\tilde{c}}^\dag \mathbf{K}_U \mathbf{\tilde{c}}=\sum_{i}\alpha\eta(s_{i,\tau})\left[\tilde{\rho}_{i\uparrow}(\tau)+\tilde{\rho}_{i\downarrow}(\tau)\right]$.
$\mathbf{B}_\tau$ matrix is defined as $\mathbf{B}_\tau=e^{\mathbf{K}_U}e^{\mathbf{K}_V}e^{\mathbf{K}_t}$.
To obtain Eq.~\eqref{eq:trace_fermion}, one has to notice that
for fermion bilinears, the identity $\text{Tr}\left[e^{\mathbf{c}^\dag\mathbf{K}_1\mathbf{c}} e^{\mathbf{c}^\dag\mathbf{K}_2\mathbf{c}} \cdots e^{\mathbf{c}^\dag\mathbf{K}_n\mathbf{c}}\right]=\text{det}\left(\mathbf{1}+e^{\mathbf{K}_1}e^{\mathbf{K}_2}\cdots e^{\mathbf{K}_n}\right)$ holds.

To summarize the result, we identify the field configurations $\{\theta,s\}$ as $\mathcal{C}$,
and the partition function can finally be expressed as
$\mathcal{Z}=\sum_{\mathcal{C}}\omega_{\mathcal{C}}^{I}\omega_{\mathcal{C}}^{II}$ with
\begin{subequations}\begin{align}
    \omega_{\mathcal{C}}^{I} &= e^{\frac{1}{2\Delta\tau K}\sum_{\left\langle ij\right\rangle,\tau}\cos\left(\theta_{ij,\tau+1}-\theta_{ij,\tau}\right) + \Delta\tau J\sum_{\langle ijk\rangle,\tau}\cos\left(\theta_{ij,\tau}-\theta_{jk,\tau}\right)} \left(\prod_{i,\tau}\gamma(s_{i,\tau})e^{-\alpha\eta(s_{i,\tau})}\right),\\
    \omega_{\mathcal{C}}^{II} &= \text{det}\left(\mathbf{1}+\prod_{\tau}\mathbf{B}_\tau\right).\label{eq:config_wight_2}
\end{align}\end{subequations}
It is important to realize that
$\mathbf{\tilde{c}}^\dag\mathbf{K}_t\mathbf{\tilde{c}}$,
$\mathbf{\tilde{c}}^\dag\mathbf{K}_V\mathbf{\tilde{c}}$,
$\mathbf{\tilde{c}}^\dag\mathbf{K}_U\mathbf{\tilde{c}}$
stay invariant under the anti-unitary transformation
$\mathcal{U}$: $c_{i,\sigma}\to \sigma \left(-\right)^i c_{i,-\sigma}^\dag$, $\sqrt{-1}\to-\sqrt{-1}$.
This anti-unitary symmetry guarantees that
the determinant in Eq.~\eqref{eq:config_wight_2} is positive for any configuration $\mathcal{C}$,
and the sign problem does not appear~\cite{wu_sufficient_2005}.

\subsection{Monte Carlo Updating Schemes}
We adopt the local updating scheme to update the auxiliary field configuration $\mathcal{C}$.
The accepting ratio $R$ of the new configuration $\mathcal{C}^\prime$ from $\mathcal{C}$
is given by the detailed balance principle as
\begin{equation}
    R = \text{min}\left\{1, \omega_{\mathcal{C}^\prime}^{I}\omega_{\mathcal{C}^\prime}^{II}/\omega_{\mathcal{C}}^{I}\omega_{\mathcal{C}}^{II}\right\},
\end{equation}
where the ratio $\omega_{\mathcal{C}^\prime}^{I}/\omega_{\mathcal{C}}^{I}$ can be calculated trivially.
To evaluate $\omega_{\mathcal{C}^\prime}^{II}/\omega_{\mathcal{C}}^{II}$, a nice result holds
\begin{equation}\label{eq:ratio2}
    \frac{\omega_{\mathcal{C}^\prime}^{II}}{\omega_{\mathcal{C}}^{II}} = \text{det} \left[\mathbf{1}+\mathbf{\Delta}\left(\mathbf{1}-\mathbf{G}(\tau,\tau)\right)\right],
\end{equation}
where we define $\mathbf{\Delta}=\mathbf{B}^\prime_\tau\mathbf{B}_\tau^{-1}-\mathbf{1}$,
and the equal-time Green's function
$\mathbf{G}(\tau,\tau)=\left(\mathbf{1}+\mathbf{B}(\tau,0)\mathbf{B}(\beta,\tau)\right)^{-1}$
with $\mathbf{B}(\tau_2,\tau_1)=\prod_{\tau=\tau_1+1}^{\tau_2}\mathbf{B}_\tau$.
We emphasise that the Green's functions are crucial to both the Monte Carlo updates and the measurements of physical observables,
hence we shall carry them around throughout the entire simulation.
Once the new configuration $\mathcal{C}^\prime$ is accepted, the Green's function is updated accordingly as
\begin{equation}\label{eq:update_green}
    \mathbf{G}^\prime(\tau,\tau) = \mathbf{G}(\tau,\tau) \left[\mathbf{1}+\mathbf{\Delta}\left(\mathbf{1}-\mathbf{G}(\tau,\tau)\right)\right].
\end{equation}

For local updates, $\mathbf{\Delta}$ possesses a quite sparse structure:
e.g. for local update $s_{i,\tau}\to s^\prime_{i,\tau}$, it has only two non-zero diagonal elements;
and for the update of rotor $\theta_{ij,\tau}\to \theta^\prime_{ij,\tau}$,
$\mathbf{\Delta}$ also involves a limited number of elements,
which does not scale with the system size $N$
so long as checkerboard breakups are applied when $e^{\mathbf{K}_V}$ is multiplied.
Due to the sparseness of $\mathbf{\Delta}$,
the locally updated Green's function can be evaluated efficiently
according to Eq.~\eqref{eq:update_green} with the Sherman-Morrison-Woodbury formula~\cite{assaad_computational_2008}.

\begin{figure}[htbp]
    \centering
    \includegraphics[width=.9\textwidth]{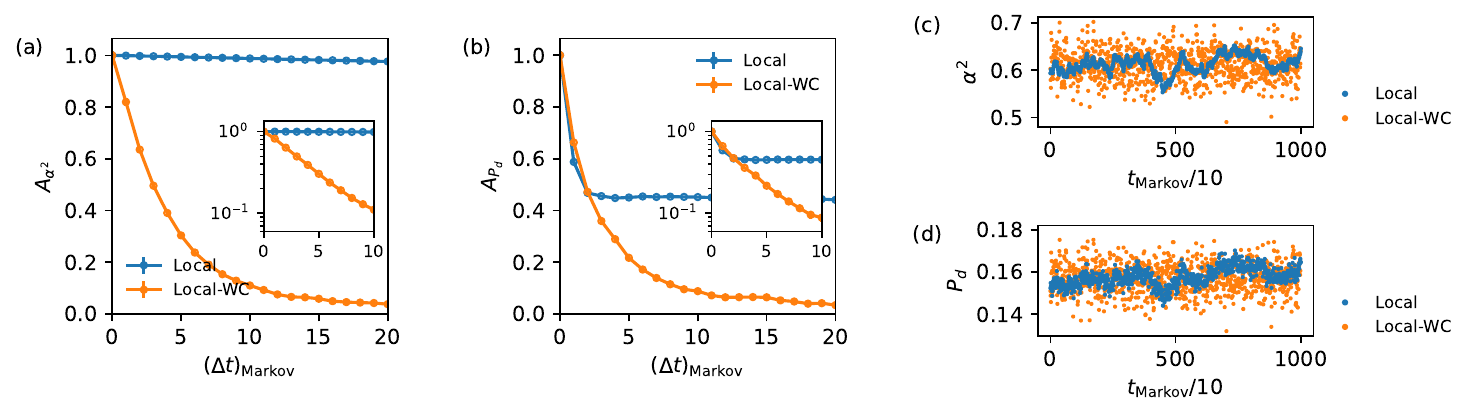}
    \caption{
        Autocorrelation functions for the local updating scheme and the combined local-WC updating scheme.
        The autocorrelation function of a certain observable $O$ is defined as
        $A_O(\Delta t)=[\langle O(t)O(t+\Delta t)\rangle - \langle O\rangle^2]/[\langle O^2\rangle - \langle O\rangle^2]$.
        $\Delta t$ denotes the time difference on the Markov chain,
        and the Markov time $t$ is measured in units of one single Monte Carlo sweep,
        where we scan over the complete space-time lattice for one time.
        In the local-WC scheme, one Wolff cluster move is followed after a local Monte Carlo sweep.
        Measurements are performed in the dSC phase with parameters $L=8$, $\beta t=2$, $\Delta\tau t=0.05$, $t=1$, $U/t=4$, $V/t=1$, and $K/t=J/t=1/2$.
        Under this setup, the local scheme typically suffers from the critical slowing down.
        (a)(b) Autocorrelation function for
               (a) the bosonic pairing correlation $\alpha^2$, and
               (b) the fermionic pairing correlation $P_d$.
               The insets show the semi-log plots of the same data
               highlighting the exponentially decaying part of the autocorrelation.
        (c)(d) Typical sequences of $\alpha^2$ and $P_d$ as function of $t_\text{Markov}$ in the equilibrium.
               Ten adjacent samples are regrouped for better visualization.
               The local-WC scheme generates nearly `randomly' distributed sequences,
               justifying the statistical independence between samples.
               In contrast, the sequences generated by the local scheme show persistent correlations
               extending over thousands of Monte Carlo sweeps for both $\alpha^2$ and $P_d$.
    }
    \label{fig:figs4}
\end{figure}

Apart from the local updates, we also develop a global updating scheme to update the rotors by cluster.
Because the rotor fields serve as the order parameter
concerned with the $\mathbb{U}(1)$ symmetry breaking of superconductivity,
the simulation under the local updating scheme suffers from the critical slowing down,
especially in the dSC phase at low temperatures with a relatively small $\Delta\tau$.
To overcome this problem, we embed the Wolff algorithm~\cite{wolff_collective_1989,jiang_solving_2022} into the general framework of DQMC.
The steps for the global updating scheme are sketched as follows:
\begin{itemize}
    \item Randomly select certain rotor field $\theta_{ij,\tau}$ as the starting point for growing the cluster.
    \item Construct the rotor cluster in both the spatial and temporal directions according to the rotor-relevant weight in $\omega_{\mathcal{C}}^{I}$,
          using the standard Wolff algorithm~\cite{wolff_collective_1989}.
    \item To fulfill the detailed balance, the trial cluster update is accepted based on a probability given by
          \begin{equation}
              \mathcal{A}_f(\mathcal{C}\to\mathcal{C^\prime}) = \text{min}\left\{1, \omega_{\mathcal{C}^\prime}^{II}/\omega_{\mathcal{C}}^{II}\right\},
          \end{equation}
          where $\mathcal{C^\prime}$ denotes the new configuration after the cluster update.
          Once the trial configuration is accepted, we update all the rotors in the cluster simultaneously,
          and this completes a Wolff cluster update.
\end{itemize}
Note that after each global update,
we need to compute the updated Green's function from scratch, which is numerically expensive.
In practice, we always combine the global updates with the local ones,
e.g. performing a batch of global updates after several local Monte Carlo sweeps over the complete space-time lattice.
It is found that in the $d$-wave superconducting (dSC) phase,
the combined updating scheme with both local and Wolff cluster (WC) updates
tremendously reduces the Markov time for the auxiliary fields reaching thermal equilibrium,
and in addition exhibits a much shorter autocorrelation time for various observables
as compared with the pure local updating scheme.
In Fig.~\ref{fig:figs4} we show the autocorrelation functions measured for representative observables with these two updating schemes.

\subsection{Measurements}~\label{sec:measurements}
Both static and dynamic fermionic observables can be measured in DQMC by decomposing them into equal-time or time-displaced Green's functions according to Wick's theorem.
A standard Monte Carlo sampling procedure then follows to estimate the expectation values of these observables.
For our purpose of detecting the $d$-wave superconductivity, we define the $d$-wave pairing order parameter $\Delta_d=\frac{1}{2N}\sum_{\langle ij\rangle}\tau_{ij}(c_{j\downarrow}c_{i\uparrow}-c_{j\uparrow}c_{i\downarrow})$,
and measure the static $d$-wave pairing correlation function $P_d$,
\begin{equation}\begin{aligned}\label{eq:dwave_pairing_corelation}
    P_d &\overset{!}{=} \Delta_d{\Delta_d}^\dag + {\Delta_d}^\dag\Delta_d\\
        &= \frac{1}{(2N)^2} \sum_{\langle ij\rangle, \langle kl\rangle} \tau_{ij}\tau_{kl} (-)^j(-)^l
        \left[\left(\tilde{c}_{j\downarrow}^\dag\tilde{c}_{i\uparrow}+\tilde{c}_{j\uparrow}\tilde{c}_{i\downarrow}^\dag\right)
        \left(\tilde{c}_{k\uparrow}^\dag\tilde{c}_{l\downarrow}+\tilde{c}_{k\downarrow}\tilde{c}_{l\uparrow}^\dag\right)
        +\left(\tilde{c}_{k\uparrow}^\dag\tilde{c}_{l\downarrow}+\tilde{c}_{k\downarrow}\tilde{c}_{l\uparrow}^\dag\right)
        \left(\tilde{c}_{j\downarrow}^\dag\tilde{c}_{i\uparrow}+\tilde{c}_{j\uparrow}\tilde{c}_{i\downarrow}^\dag\right)\right],
\end{aligned}\end{equation}
where we have expressed the formula in the particle-hole channel, and identified the equal-time Green's function in this basis as $\mathbf{G}_{i\sigma,j\sigma^\prime}(\tau,\tau)\overset{!}{=}\langle\tilde{c}_{i\sigma}(\tau)\tilde{c}_{j\sigma^\prime}^\dag(\tau)\rangle$.
One then breaks up the quartic terms in Eq.~\eqref{eq:dwave_pairing_corelation} into products of Green's function using Wick's theorem.

Another important observable we desire to evaluate is the single-particle spectral function $A_\sigma(\bm{k},\omega)$,
through the measurement of which the formation of pseudogap and Fermi arcs are directly observed.
Consider the time-displaced Green's function in the momentum space,
\begin{equation}\begin{aligned}
    G_{\sigma}(\bm{k},\tau) &\overset{!}{=} \frac{1}{N} \sum_{ij} e^{-i\bm{k}\left(\bm{r}_i-\bm{r}_j\right)} \left\langle c_{i\sigma}(\tau) c^\dag_{j\sigma}(0) \right\rangle\\
    &= \left\{
        \begin{array}{lr}
            \frac{1}{N} \sum_{ij} e^{-i\bm{k}\left(\bm{r}_i-\bm{r}_j\right)} \mathbf{G}_{i\sigma,j\sigma}(\tau,0), & \quad\sigma=\ \uparrow\\[5pt]
            -\frac{1}{N} \sum_{ij} e^{-i\bm{k}\left(\bm{r}_i-\bm{r}_j\right)} (-)^i(-)^j \mathbf{G}_{j\sigma,i\sigma}(0,\tau), & \quad\sigma=\ \downarrow\\
        \end{array}
       \right.
\end{aligned}\end{equation}
where
$
\mathbf{G}_{i\sigma,j\sigma^\prime}(\tau,0) \overset{!}{=} \langle \tilde{c}_{i\sigma}(\tau)\tilde{c}_{j\sigma^\prime}^\dag(0) \rangle
$
and
$
\mathbf{G}_{i\sigma,j\sigma^\prime}(0,\tau) \overset{!}{=} \langle \tilde{c}_{i\sigma}(0)\tilde{c}_{j\sigma^\prime}^\dag(\tau) \rangle = -\langle \tilde{c}_{j\sigma^\prime}^\dag(\tau)\tilde{c}_{i\sigma}(0) \rangle
$.
The imaginary-time Green's function $G_{\sigma}(\bm{k},\tau)$ is directly measurable in DQMC,
while $A_\sigma(\bm{k},\omega)$ is not.
In fact, the single-particle spectrum $A_\sigma(\bm{k},\omega)$
is related to $G_{\sigma}(\bm{k},\tau)$ through an integral transform,
for obtaining which we perform the stochastic analytic continuation calculations,
as we will introduce below in Sec.~\ref{sec:sac}.
%--------------------------------------------------------------------------

\section{Absence of Antiferromagnetic Order}~\label{sec:no-afm}
\begin{figure}[htbp]
    \centering\hspace*{1.5cm}
    \includegraphics[width=0.75\textwidth]{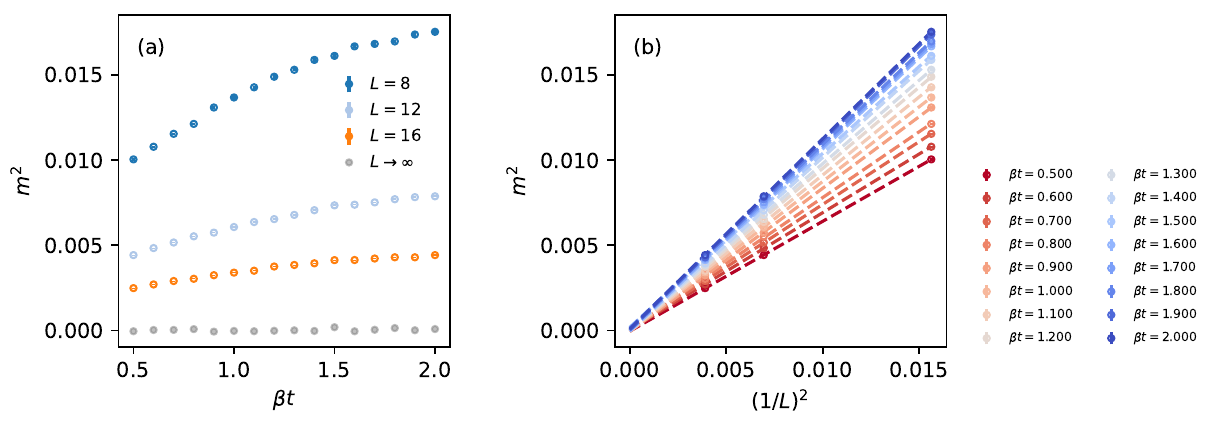}
    \caption{
        Suppression of static AFM correlation $m^2$.
    }
    \label{fig:figs5}
\end{figure}

%--------------------------------------------------------------------------
It is shown in this section that
in our explored parameter region of the Hubbard-like model,
there is vanishing antiferromagnetic order (AFM) at finite temperatures
although strong on-site electron repulsion is present.
The AFM order parameter is defined as $\vec{m} = \frac{1}{N} \sum_i (-)^i \vec{S}_i$,
with the static correlation
\begin{equation}
    m^2 = \frac{1}{N^2} \sum_{ij} (-)^i (-)^j \left\langle \vec{S}_i \cdot \vec{S}_j \right\rangle.
\end{equation}
$\vec{S}_i=\frac{1}{2} \sum_{\alpha\beta} \vec{\sigma}_{\alpha\beta} c^\dag_{i\alpha} c_{i\beta}$ denotes the electron spin at site $i$,
and $\sigma^{a}$ are three Pauli matrices for $a=x,y,z$.
The spin correlation is then expressed in terms of fermion operators, 

\begin{equation}\begin{aligned}
    \vec{S}_i \cdot \vec{S}_j
    &= \frac{1}{4} (\rho_{i\uparrow}-\rho_{i\downarrow}) (\rho_{j\uparrow}-\rho_{j\downarrow})
     + \frac{1}{2} \sum_{\alpha} c^\dag_{i\alpha}c_{i\bar{\alpha}}c^\dag_{j\bar{\alpha}}c_{j\alpha}\\[4pt]
    &= \frac{1}{4} (\tilde{\rho}_{i\uparrow}+\tilde{\rho}_{i\downarrow}-1) (\tilde{\rho}_{j\uparrow}+\tilde{\rho}_{j\downarrow}-1)
     + \frac{1}{2} (-)^i(-)^j \left(\tilde{c}^\dag_{i\uparrow}\tilde{c}^\dag_{i\downarrow}\tilde{c}_{j\downarrow}\tilde{c}_{j\uparrow} + \tilde{c}_{i\downarrow}\tilde{c}_{i\uparrow}\tilde{c}^\dag_{j\uparrow}\tilde{c}^\dag_{j\downarrow}\right),
\end{aligned}\end{equation}
where the two-body correlations involved are now ready to be broken down into products of one-body Green's functions in the particle-hole basis.
In Fig.~\ref{fig:figs5}, we measure the static AFM correlation $m^2$ as a function of $\beta$ and $L$.
As adopted in the main text, we have set $t=1$, $U/t=4$, $V/t=1$, $K/t=J/t=1/2$, and $\Delta\tau=0.025/t$.
The examined temperature range covers both the dSC phase and the normal state.
In Fig.~\ref{fig:figs5}(b), the value of $m^2$ in the thermodynamic limit is estimated by extrapolating $m^2(L)$ to $1/L\to0$,
which proves the absence of AFM order.

%--------------------------------------------------------------------------

\section{Determination of the BKT Transition Point}
%--------------------------------------------------------------------------
The precise determination of the BKT transition point requires careful finite-size scaling analysis.
According to the finite-size-scaling theory~\cite{challa_critical_1986},
the scaling of the $d$-wave pairing correlation $P_d$ follows a BKT form
\begin{equation}\label{eq:finite_size_scaling}
    P_d L^{\eta(T_c)} = f\left[\exp(b t_r^{-1/2})/L\right], \quad \xi \sim \exp\left(b t_r^{-1/2}\right).
\end{equation}
$t_r=T/T_c-1$ denotes the reduced temperature, and $f$ is a universal scaling function.
The critical exponent is $\eta(T_c)=1/4$ for the BKT transition~\cite{kosterlitz_critical_1974}.
For the bosonic correlation $\alpha^2$, an identical relation as in Eq.~\eqref{eq:finite_size_scaling} applies.
At $T=T_c$ where the correlation length $\xi$ diverges,
it is inferred from Eq.~\eqref{eq:finite_size_scaling} that $P_d L^{\eta(T_c)}$ for varying $L$ shall coincide at $T_c$.
In DQMC simulations, the $d$-wave pairing correlation $P_d(L,\beta)$ is measured for varying temperatures and system sizes.
It is thus expected that $P_d(L,\beta) L^{\eta(T_c)}$ converge at $\beta_c$ for all $L$,
which allows us to precisely locate the transition temperature $\beta_c$,
as we have done in the main text.
Given the $\beta_c$ obtained above,
one then tunes the non-universal factor $b$ to
let the $P_d(L,\beta)$ data for all $\beta$ and $L$ collapse into a universal function
according to Eq.~\eqref{eq:finite_size_scaling}.
Once the value of $b$ is determined, the correlation length $\xi$ follows as $\xi\sim\exp(b t_r^{-1/2})$.
%--------------------------------------------------------------------------

\section{Stochastic Analytic Continuation Calculation}~\label{sec:sac}
%--------------------------------------------------------------------------
As mentioned in Sec.~\ref{sec:measurements},
it is a technically challenging task to 
evaluate the single-particle spectrum $A(\omega)$ from the imaginary-time Green's function $G(\tau)$,
which is defined on a set of discrete imaginary-time points $\{\tau\}$.
By definition, $G(\tau)$ is related to $A(\omega)$ through the following integral transform
\begin{equation}
    G(\tau) = \int \mathrm{d}\omega A(\omega) K(\tau,\omega),
\end{equation}
where $K(\tau,\omega)$ denotes the integral kernel.
For fermionic Green's function, $K(\tau,\omega)=e^{-\tau\omega}/\left(1+e^{-\beta\omega}\right)$.
It has long been realized that solving the inverse integral transform, i.e. obtaining $A(\omega)$ from $G(\tau)$,
appears to be an ill-posed problem.
Practically, $G(\tau)$'s are always estimated from the QMC simulations with certain statistical errors,
which makes the analytic continuation procedure numerically unstable.

To overcome this challenge,
we apply the stochastic analytic continuation (SAC)~\cite{sandvik_stochastic_1998} algorithm
to obtain all the single-particle spectra presented in this work.
The combination of QMC and SAC has been shown to reliably
reconstruct the real-frequency spectral functions among a wide class of spin and fermionic models.
For interested readers, we refer to a recent review~\cite{shao_progress_2023},
which provides an in-depth introduction to SAC.
The key spirit of SAC is to perform a Monte Carlo sampling procedure over the parametric space of $A(\omega)$.
The statistical weight of certain parametric configuration $\mathcal{C}_A$ is assumed to obey a Boltzmann distribution,
\begin{equation}
    W_{\mathcal{C}_A} \sim \exp \left( -\frac{\chi^2}{2\Theta} \right),
\end{equation}
where $\chi^2$ quantifies the goodness of fitting between the QMC-measured Green's functions and the ones computed from the parameterized spectrum.
$\Theta$ represents the fictitious temperature
that serves to balance the minimization of $\chi^2$ and the `thermal' fluctuations, which help to smooth the spectrum.
In practice, one performs a simulated annealing procedure to lower $\Theta$ adiabatically from a sufficiently high value.
Once the optimal $\Theta$ is found according to certain criteria,
all statistically equivalent spectral configurations are sampled to produce the final real-frequency spectrum.
%--------------------------------------------------------------------------

%--------------------------------------------------------------------------
\end{document}